\documentclass[useAMS,usenatbib,usegraphicx]{mn2e}
\usepackage{amsmath}

\newcommand{\Mo}{\rm\thinspace M$_\odot$\,}
\newcommand{\rg}{\rm\thinspace $r_{\rm g}$\,}

\title[Iron K Reverberation in NGC 4151]{Relativistic iron K X-ray Reverberation in NGC 4151}
\author[A. Zoghbi et al.]{A. Zoghbi$^{1,2}$\thanks{E-mail:
azoghbi@astro.umd.edu}, A. C. Fabian$^{2}$, C. S. Reynolds$^{1}$ and E. M. Cackett$^{2,3}$\\
$^{1}$Department of Astronomy, University of Maryland, College Park, MD 20742, USA\\
$^{2}$Institute of Astronomy, Madingley Road, Cambridge CB3 0HA\\
$^{3}$Department of Physics and Astronomy, Wayne State University,  Detroit, MI 48201, USA}

\begin{document}

\date{}

\pagerange{\pageref{firstpage}--\pageref{lastpage}} \pubyear{2011}

\maketitle

\label{firstpage}

\begin{abstract}
Recent X-ray observations have enabled the study of reverberation delays in AGN for the first
time. All the detections so far are in sources with a strong soft excess, and the measured delay is
between the hard (1--3 keV) direct continuum and the soft excess (0.5--1 keV), interpreted as the reflection continuum 
smeared by relativistic effects. There is however an inherent ambiguity in identifying and studying
the details of the lines in the soft excess. Here we report the first detection of reverberation in the iron K band in any AGN. Using XMM-Newton observations of NGC~4151, we find delays of order 2000 s on time-scales of $10^5$ s between the $5-6$ keV band and $2-3$ and $7-8$ keV bands, with a broad lag profile resembling a relativistically-broadened iron line. The peak of the lag spectra shifts to lower energies at higher frequencies, consistent with the red wing of the line being emitted at smaller radii, as expected from reflection off the inner accretion disk. This is a first detection of a broad iron line using timing studies.
\end{abstract}


\section{Introduction}
X-ray observations of active galactic nuclei (AGN) have shown that most of the radiation is emitted from very close to the central super-massive black hole. Broadening of X-ray spectral features caused by relativistic effects has been the main tool in studying these systems (\citealt{2007ARA&A..45..441M} for a review).
Recent work has moved beyond spectral modelling to detecting time delays between the different spectral components. Short time delays have first been seen in the Narrow Line Seyfert 1 galaxy 1H0707-495 (\citealt{2009Natur.459..540F,2010MNRAS.401.2419Z}). Thanks to its strong soft excess and high iron abundance, small delays of a few tens of seconds are detected between the 1--4 and 0.5--1 keV bands where the soft band lags the hard band (hence soft lag). The natural explanation for the lag is reprocessing by the reflecting accretion disc in the vicinity of the black hole. Similar delays have also been seen in several other objects (\citealt{2011MNRAS.416L..94E,2011MNRAS.417L..98D,2011MNRAS.418.2642Z}). A common feature among all these sources is a strong soft excess in the spectrum below 1 keV. This is interpreted, in line with the lags, as a relativistically-blurred reflection continuum, that responds to variations in a primary illuminating hard power-law.

The interpretation of the soft lag (soft excess lagging the hard continuum) as a light crossing effect close to the black hole was questioned by \cite{2010MNRAS.408.1928M}, who suggested that such small delays could be an artefact of a much larger reverberating system of clouds ($\sim$1000 instead of $\sim$ a few gravitational radii \rg). Although this interpretation fails when a complete set of
observables (variability power, flux and lag spectra) are considered (\citealt{2011MNRAS.412...59Z}), it stands on the ambiguity in interpreting the soft excess itself (e.g. \citealt{2006MNRAS.365.1067C,2011arXiv1107.5429D}). The smoothness of its spectrum makes identifying the relativistically-broadened emission lines therein difficult unless the object has some extreme properties (e.g. high iron abundance and strong Fe-L in 1H0707-495). Therefore, a key challenge for any model is its ability to explain the reverberation in the iron K region (6--7 keV) where such an ambiguity is not an issue. The vital task is to find and observe an object that has reverberation signatures in this band. The importance of reverberation studies in the K band have been discussed previously, along with theoretical calculations of the expected behaviour of the broad iron line with time (\citealt{1990Natur.344..747S,1992MNRAS.259..433M,1999ApJ...514..164R}). However, it remained difficult to detect these reverberation delays in real data.

The count rate at the bands of interest is a key parameter in reverberation studies. All previous lag detections have been in the soft band ($< 1$ keV) in objects with a strong soft excess. Those objects however, show steep spectra where the flux drops significantly in the iron-K band. Therefore, a different class of objects need to be explored if Fe-K reverberation is to be studied. In this work we show that NGC~4151, which is $\sim300$ times brighter than 1H0707-495 at 6.4 keV, shows inter-band delays that indicate that a significant fraction of the flux in this band originates in the vicinity of the black hole.

NGC 4151 is a well-studied Seyfert galaxy ($z=0.0033$) that harbours one of the brightest AGN in the X-ray band. It has been observed and studied by all major X-ray satellites. The X-ray to $\gamma$-ray spectrum is dominated by a Comptonisation component that breaks at $\sim 100$ keV. There is a strong narrow iron line at 6.4 keV and a reflection hump peaking at 30 keV (\citealt{1996MNRAS.283..193Z,2002MNRAS.334..811S}). These components are absorbed below 4 keV by a high column ($N_H\sim10^{22} {\rm cm}^{-2}$) of gas in the line of sight (\citealt{1994ApJ...436L..27W}). The spectrum below 2 keV is very complex. High resolution Chandra observations revealed that it is spatially-resolved and it is associated with a highly ionised plasma coincident with the optical narrow-line region (\citealt{2000ApJ...545L..81O,2011ApJ...729...75W}). This component does not vary much and will not be discussed here.

\begin{figure}
\centering
 \includegraphics[width=210pt,clip ]{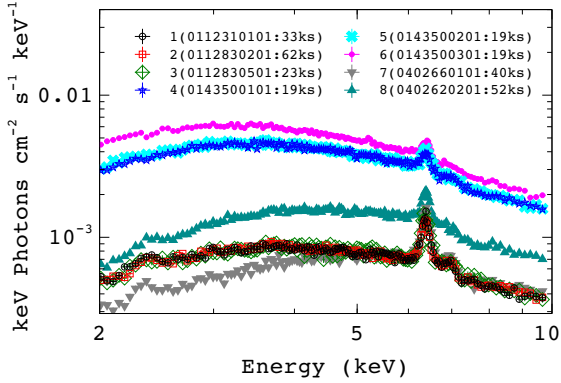}
\caption[Unfolded spectra of NGC 4151]{Unfolded spectra (in $EF(E)$ form) of NGC 4151 from the XMM-Newton PN detector. The labels include: Obs. number (Obs. ID : total exposure time). }
\label{fig:unfold_spec}
\end{figure}

NGC 4151 has been observed with XMM-Newton several times in short observations constrained by its visibility. The unfolded spectra are shown in Fig. \ref{fig:unfold_spec} for the eight longest observations ($\sim19-62$ ks) from a total of 13 XMM-Newton. There is clearly significant spectral variability. The difference between the spectra can be as simple as a uniform apparent increase in flux, or it could be due to changes in the absorber on time scales of years (e.g. difference between observations 1 and 7 in Fig. \ref{fig:unfold_spec}).

In this work, light curves from eight archival XMM observations are used (Obs. IDs and length of observations are shown in Fig. \ref{fig:unfold_spec}). ODF data were reduced using \textsc{SAS v. 11.0.0}. Light curves are produced in equal energy steps above 2 keV with 1024 s time bins. Instrumental corrections were applied using the task \texttt{epiclccorr}. The lag was then estimated at the observed times following \cite{1992ApJ...398..169R}. We determine the uncertainty in the lags via Monte Carlo simulations.  We shift each point in the light curves by drawing a random number from a Gaussian distribution with mean equal to the count rate and standard deviation equal to the uncertainty in the count rate.  This is performed 100 times, and the 1-sigma uncertainty in the lag is taken from the resulting distribution. The errors are very similar to those estimated in the standard way (\citealt{1999ApJ...510..874N}). Lag-energy plots are produced by calculating the lag at the frequency of interest between the energy band and the light curve in the whole 2--10 keV band excluding the current band. This maximises the signal to noise and essentially measures the delays relative to the average (e.g. \citealt{2011MNRAS.412...59Z}).

\section{Lag measurement in NGC 4151}\label{lag_measure}
\begin{figure}
\centering
 \includegraphics[width=220pt,clip ]{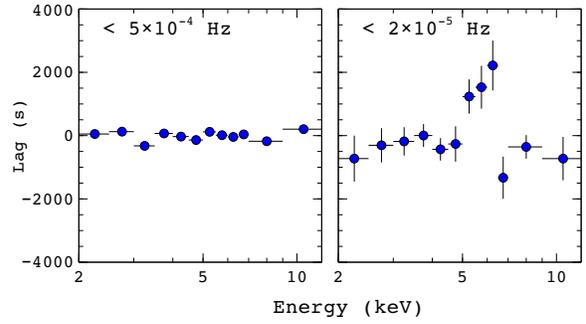}
\caption{{\it Left:} Lag spectrum at frequencies smaller than $5\times10^{-4}$ Hz for NGC 4151 using all the eight observations. The delays are measured with respect to the total 2--10 keV band (i.e. the lags of all the points are measured with respect to roughly their average). {\it Right:} Same as {\it Left} but only for the lowest frequencies.}
\label{fig:lag_all}
\end{figure}

The mass of NGC 4151 is determined accurately with H$\beta$ reverberation mapping to be $M=4.5\pm0.5\times10^7$ \Mo (\citealt{2006ApJ...651..775B}). If the time scales seen in 1H0707-495 and RE J1034+396 are scaled with mass, reverberation delays are expected at $\sim3\times10^{-5}$ Hz. Fig. \ref{fig:lag_all}-left shows the lag spectrum for all the observations measured at frequencies $<5\times10^{-4}$ Hz, calculated in a similar way to those in \cite{2011MNRAS.412...59Z}.

In the right panel of Fig. \ref{fig:lag_all}, we show the lag spectra for the lowest frequencies (derived from a single frequency bin estimate) as they are estimated from the longest observation (i.e. the 2nd shown in Fig. \ref{fig:unfold_spec}). 
This plot clearly shows that the $\sim6$ keV band lags the $>8$ and $<4$ keV bands by $\sim 2000$ seconds on time scales of $\sim 10^5$ seconds. Since this observation happens to map a low flux source state this could be a possible indication of a relation between the lag and the flux state.

\begin{figure}
\centering
 \includegraphics[width=130pt,clip ]{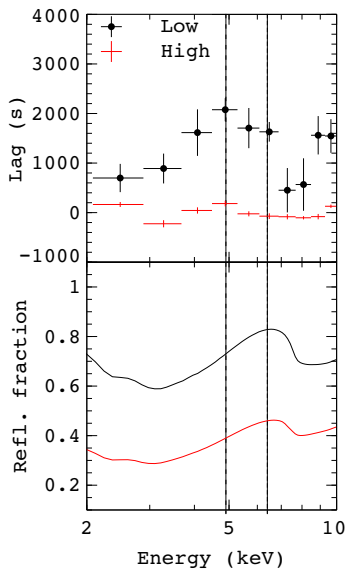}
\caption{{\it Top:} Lag spectrum at frequencies $(0.5-5)\times10^{-4}$ Hz for NGC 4151 for the low and high flux states separately. The plots are similar to Fig. \ref{fig:lag_all}. Lags for the low flux have been shifted up by 1500 seconds for visual clarity. {\it Bottom:} Reflection fraction calculated as the ratio of the relativistically blurred component and the power-law component taken from the best fit to observations 1 and 4 representing low and high flux intervals respectively (see Sec. \ref{lag_interp}). Vertical lines mark the peaks in both the lag and flux spectra (at 4.9 and 6.4 keV for the lag spectrum and reflection fraction respectively).}
\label{fig:lag_low_hi}
\end{figure}

Calculating the lags for the low and high flux intervals separately shows that the lags are prominent only in the low flux observations. To avoid the obvious effects of changes due to the absorber, low flux observations include observation 1,2 and 3 (as defined in Fig. \ref{fig:unfold_spec}). The high flux observations are 4, 5 and 6. Fig. \ref{fig:lag_low_hi}-top shows the lag-energy plot for the two flux intervals calculated between $(0.5-5)\times10^{-4}$ Hz. The lower limit is dictated by the length of the high flux observations. The conclusions are not very sensitive to the upper limits apart from a scaling factor for the lag and slightly larger errors if smaller band is considered. Again, at low fluxes, the $\sim5$ keV band lags both the $2-3$ and $8-9$ keV bands, while the lag spectrum is significantly different at high fluxes. The plot clearly shows an interesting structure at the iron K energies that resembles a broad iron line.  It should be noted here that the lags are not very prominent when all observations are included because the high-flux signal dominates when combining the data. It should be noted also that the peak in the lag spectrum measured at $(0.5-5)\times10^{-4}$ Hz is at lower energies compared to that measured at lower frequencies (Fig. \ref{fig:lag_low_hi}-top vs. \ref{fig:lag_all}), an effect which is discussed in Sec \ref{lag_interp}. 

\begin{figure}
\centering
 \includegraphics[width=220pt,clip ]{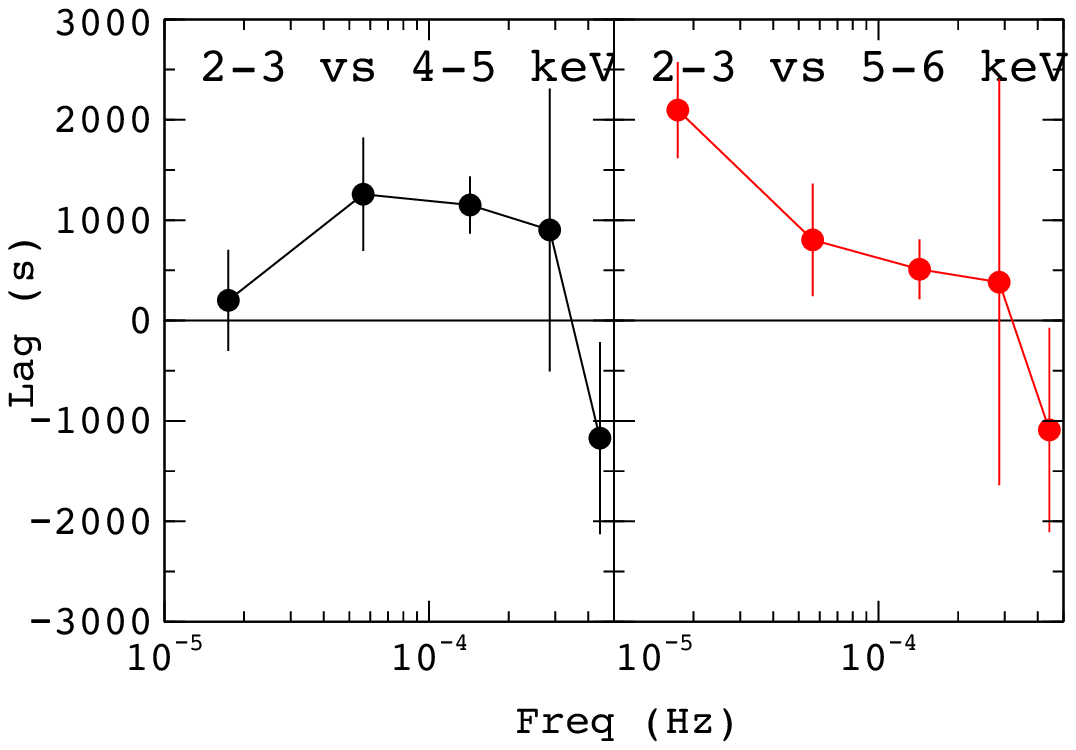}
\caption{Lag vs frequency plot for the low flux interval. Hard lags are plotted, so a positive delay indicate hard bands lagging softer bands. {\it Left:} The lag between 2--3 and 4--5 keV bands. {\it Right:} The lag between 2--3 and 5--6 keV. The energy and frequency dependence of the lag is apparent.}
\label{fig:lag_vs_fq}
\end{figure}

The frequency-dependent lag is shown in Fig. \ref{fig:lag_vs_fq} for the low flux observations. It shows how the lag depends on energy as well as frequency. The lag peaks at lower frequencies when higher energies are considered (Fig. \ref{fig:lag_vs_fq}-right), and the peak is at higher frequencies when a lower energy band is considered (Fig. \ref{fig:lag_vs_fq}-left). Due to the length of the observations, the lowest frequency point contains one frequency value (from the longest observation). This can be problematic because of the small number of points included in the averaging process and only longer future observations will provide any improvements. There are however several reasons that strengthen our measurements of an energy and frequency dependence. As shown in \cite{2010MNRAS.401.2419Z}, Monte Carlo simulations show that the lowest frequency point always shows a bias toward zero and the bias is proportional to the absolute value of the lag (when measured at the same frequency). Therefore, because the difference in value of the lowest frequency point between the left and right of Fig. \ref{fig:lag_vs_fq} is significant, their values might be biased but their difference has to be intrinsic. This is further supported by a clearer difference at the two higher frequency points, where the bias is irrelevant.


\section{Lag interpretation.}\label{lag_interp}
To understand these lag spectra, we turn to the flux spectra. It was realised from early observations that the source is seen through a high column density absorber. However, a model consisting of a highly absorbed power-law is not sufficient to fit the spectral curvature in the 2--5 keV band (\citealt{1994ApJ...436L..27W}). Instead, it can be reproduced by adding a partially ionised (warm) absorber, another neutral absorber, or a combination where some absorbers only partially cover the source (\citealt{1994ApJ...436L..27W,1995MNRAS.275.1003W,2002MNRAS.334..811S,2007MNRAS.377..607P}). However, in a systematic analysis of the Fe K band for a sample of AGN, \cite{2007MNRAS.382..194N} noted that there is red wing to the line that cannot be accounted for by the absorbers. A similar conclusion was reached earlier by \cite{1995ApJ...453L..81Y}.

The lags shown in the top of Fig. \ref{fig:lag_low_hi} are clear signatures of reverberation due to reflection. The width of the line-like feature in the lag spectrum does not necessarily mean that the actual spectral line is broad. A distant reflector with a narrow line at 6.4 keV and an edge at 7.1 keV can possibly produce a broad line-like structure similar to that seen here (e.g. \citealt{2001MNRAS.327..799K} for black hole binaries modelling). Therefore, could this be due to the cold reflection component, which is responsible for the strong narrow line at 6.4 keV in the flux spectrum (Fig. \ref{fig:unfold_spec})?

There are several points that can help answer the question. The change in the lag spectrum between the low and high flux states is dramatic. Fig. \ref{fig:lag_low_hi}-top indicates that the lag is scaled by a factor of $\sim20$, while the source flux changes by a factor of $\sim4-5$ (Fig. \ref{fig:unfold_spec}). The flux of the narrow line in the flux spectrum does not change significantly with flux (\citealt{2003MNRAS.345..423S}), so if the lags are produced in the narrow line, the lag change with flux has to be due a change in the illuminating source and not in the reflecting medium or its geometry. In other words, a constant narrow line with flux cannot be responsible for a variable lag spectrum.

\begin{figure}
\centering
\includegraphics[width=200pt,clip ]{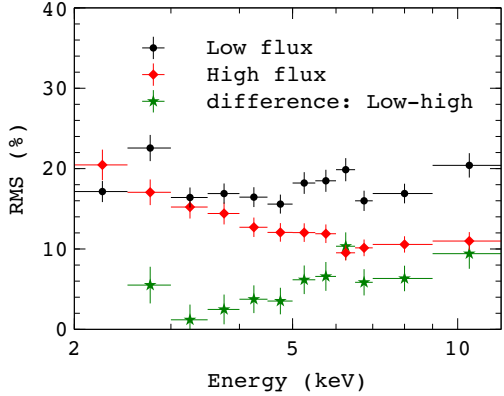}
\caption{The RMS spectrum calculated for the low and high flux observations. The difference resembles a reflection spectrum.}
\label{fig:rms_spec}
\end{figure}
Also, the value of the lag is too small to originate in the cold distant reflector. Early observations of the narrow line variability constrained it to be at least a few light years away from the central source (e.g. \citealt{2003ApJ...598..935M,2003MNRAS.345..423S}). Smaller lags can be produced by a distant reflector if the reflecting medium is patchy (\citealt{2010MNRAS.403..196M}). However, the difficulty there is that the lag has an oscillatory behaviour that makes the lag measured over a wide frequency range, as is the case here, averages to zero. The alternative of using an energy-dependent transfer function could produce small lags over a wide frequency range but it has to be finely tuned by putting the reflecting medium exactly in our sight-line to produce the observations (\citealt{2011MNRAS.412...59Z}).

The lag spectrum in Fig. \ref{fig:lag_low_hi} has a wing that extends below 6.4 keV, which indicates a possible origin in a red-shifted line. Whichever component is producing the lag, it appears to be present (or more prominent) in the low flux spectrum and not in the high flux. Taking the difference spectrum between the two states (e.g. observations 4 and 1 in Fig. \ref{fig:unfold_spec}), we find that an absorbed power-law model is not sufficient, the data requires an additional component that matches the shape of the lag spectrum and resembles a broad iron line. The RMS spectrum (\citealt{2002ApJ...568..610E}) for the low and high intervals is also different (Fig. \ref{fig:rms_spec}). At high fluxes, the RMS spectrum is decreasing smoothly with a drop at the 6.4 keV, while there is extra variability between 4--6 keV at low fluxes. The shape of this extra variability component resembles a reflection spectrum. The RMS spectrum further confirms that there is a reflection component that is more prominent in the low flux observations.

\begin{figure}
\centering
 \includegraphics[width=230pt,clip ]{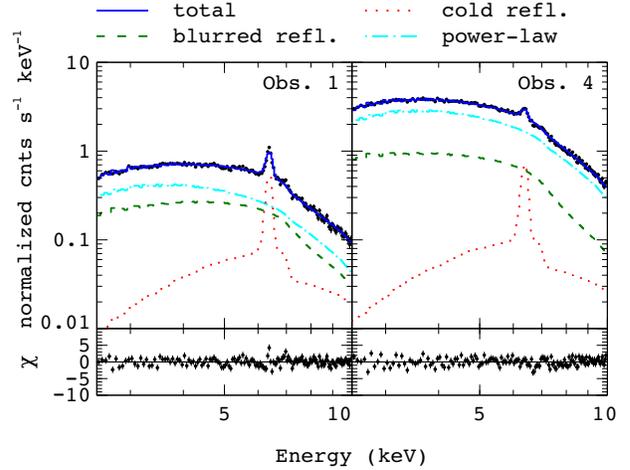}
\caption{The spectra of NGC 4151 in the low and high states (observations 1 and 4 in Fig. \ref{fig:unfold_spec} respectively), with the best fitting model. The continuous line is the total model. The cyan dot-dashed line is a power-law, the red dotted line is a cold reflector, and the green dashed line is a relativistically blurred reflection model. The residuals of the fit are plotted in the bottom.}
\label{fig:best_fit}
\end{figure}

In order to further understand the differences between the low and high flux intervals, we fitted the spectra with an absorbed power-law plus cold reflection modelled with the reflection table \emph{reflionx} (\citealt{2005MNRAS.358..211R}). Motivated by the shape of the lag spectrum, we also included a relativistically-blurred reflection component. Again we used \emph{reflionx} to model the reflection spectrum, this time, convolved with relativistic kernel \texttt{kdblur} to account for the blurring due to special and general relativistic effects. Fig. \ref{fig:best_fit} shows the best fit for the low and high flux intervals (plotted for observations 1 and 4 as an example). In the high flux interval, the spectrum consists of an absorbed power-law ($N_h=5\times10^{22} \rm{cm}^{-2}$), a cold distant reflector (dotted line) and relativistically-blurred reflection (dashed). At low fluxes, the relative contribution of the blurred reflection is higher. In other words, the main difference between the two flux observations is a difference in the reflection fraction from the blurred component. The reflection fraction is plotted in Fig. \ref{fig:lag_low_hi}-bottom, where it is taken as the ratio of the relativistically-blurred component to the power-law. This gives the relative contribution of each component to the different energy bands. The small residuals in Fig. \ref{fig:best_fit} around 6 keV are possibly due to the fact that we used {\texttt reflionx} to model the cold reflection. The model has a non-zero lower ionisation limit. 

Comparing the top and bottom of Fig. \ref{fig:lag_low_hi}, the absence of lag in the high flux observations is easily understood.  The observed lag has contributions from both the reverberation lag between the power-law and the reflection component as well as any lag between the power-law component and  itself in the different energy bands.  The strength of the contribution to the observed lag from each component is determined by which component is
dominating the variability.  At high fluxes, the contribution of reflection is much smaller, and thus the measured lag will be dominated by the lag between the power-law in each band (equal $\sim0$) rather than the reverberation lag.

\begin{figure}
\centering
 \includegraphics[width=260pt,clip ]{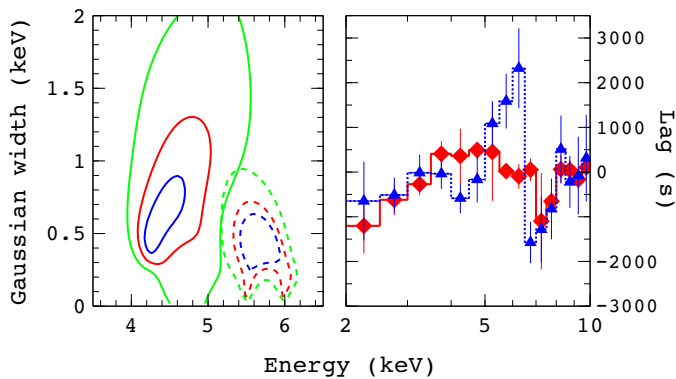}
\caption{{\it Right:} Lag-energy plots (similar to Fig. \ref{fig:lag_all},\ref{fig:lag_low_hi}). Blue triangels are for $<2\times10^{-5}$ Hz and red diamonds are for $(5-50)\times10^{-5}$ Hz. The position of zero in this panel is arbitrary. {\it left} Confidence contours for the width and energy of the gaussian lines fitted to the lag spectra. Dashed and continuous lines are for frequencies $<2\times10^{-5}$ Hz  and $(5-50)\times10^{-5}$ Hz respectively. The contours represent 90, 95 and 99 percent limits shown in blue, red and green respectively.}
\label{fig:gauss_cont}
\end{figure}

It is very interesting to note that the peak in the lag spectrum is at lower energies compared to the broad iron line (4.9 vs 6.4 keV). The two vertical lines in Fig. \ref{fig:lag_low_hi} mark the positions of the peaks in the lag and reflection fraction spectra. If only low frequencies are considered (Fig. \ref{fig:lag_all}), the peak in the lag spectrum  moves to higher energies. To explore the apparent frequency-dependence of the lag spectra we extract lag spectra at two frequency bins $<2\times10^{-5}$ and $(5-50)\times10^{-5}$ Hz (Fig. \ref{fig:gauss_cont}). Then we fitted it with a simple model consisting of a gaussian plus a constant. The two gaussians have energies $5.7\pm0.19$ and $4.60\pm0.43$ keV for the two frequency bands respectively. Confidence contours for the line energy and width are shown in Fig \ref{fig:gauss_cont}. Although there is an overlap in the width considering the 99 percent confidence contours, the energy is clearly higher at lower frequencies. This result has been tested also by calculating the lag spectrum at low frequencies, and accumulating more and more frequencies. As higher frequencies are added, the width of the line increases and the gaussian energy shifts to lower energies. This makes the result at low frequencies more robust and not just an artefact of the small number of averaged frequencies (see. end of sec. \ref{lag_measure}). Interpreting this behaviour is straightforward if the delays are due to the reverberating broad Fe-K line. Redder parts of the broad line (i.e. lower energies) originate in smaller radii and would naturally vary on smaller time-scales. Bluer parts of the line are emitted slightly further out and so vary on slightly longer time-scales.

\section{Discussion and Conclusions}
All previous lag detections in AGN have been in the soft band ($<1$ keV). In this work we reported the first detection of inter-band delays in the Fe-K band in an AGN. The lag is seen prominently in low flux observations and it is much smaller at high fluxes. The lag is in a sense that the $5-6$ keV band lags both $2-3$ and $8-9$ keV bands by $\sim2000$ seconds (this value is frequency-dependent). The interpretation of these delays is consistent, to first order, with a picture in which an illuminating power-law component, dominating at $<\sim3$ and $>7$ keV is reflected by an accretion disk at distances of $\sim 10$ \rg (The light crossing time at 1 \rg is $5\times10^{-5} M$ seconds, where $M$ is the black hole mass in solar units). The measured delays are due to light-travel time between the illuminating corona and a reflecting accretion disk. The main difference between low and high flux observations is a difference in the reflection fraction. Because the energy bands used do not contain a single component (direct power-law only or reflection only), the reduced reflection fraction causes the delays to be smaller in the high flux observations because they are diluted by the direct component where there is no clear lag between the energy bands.

The shape of the lag spectrum resembles a broad iron line peaking around 5 keV. Although the flux spectrum has a strong narrow line component, several arguments indicate that it is not the origin of the delays. First, the delays are small in value ($\sim 2000$ s), which is much smaller than the distance scales of the region producing the narrow line (the molecular torus or the broad line region). Earlier work, constrained the region producing the narrow line to be on scales light years (\cite{2000ApJ...545L..81O}). The lag spectrum peaks well below 6.4 keV, or in other words, the most lagged band is not 6.4 keV corresponding to the peak of the line as would be expected if the narrow component is producing the lag. Furthermore, the shift of the peak with frequency (Fig. \ref{fig:gauss_cont}) rules out models where the delays are due to path-difference close to the line-of-sight (e.g. \citealt{2010MNRAS.408.1928M}). In those models, no frequency-dependent behaviour is expected, as the shape of the distant reflection spectrum is not expected to change on these time-scales.

The light-crossing time-scale for NGC~4151 with a mass of $4\times10^7$ \Mo at say 6 \rg is $\sim1200$ s, which is of the same magnitude as those measured here. The thermal time-scale at this radius for a standard thin disk (\citealt{1973A&A....24..337S}), which could be the variability time-scale if ionisation/density changes are involved, is $\sim2\times10^5$ seconds assuming a viscosity parameter $\alpha$ of 0.1. Again, this time-scale is also consistent with the observations. Spectral fitting with a relativistic model gives an inner disk radius of $1.4\pm0.1$ \rg with an emissivity index of $4\pm0.3$, a profile flatter than usually observed. If the lag spectra are fitted with a relativistic line model using the inclination and emissivity from the flux spectra, we find that the low and high frequency variability originate at $\sim 10$ and $\sim4$ \rg respectively. This also presumably indicate that the height of the illuminating source is not very large (a few \rg ).

The lag spectrum of NGC~4151 consists of broad line-like feature that peaks between 4--6 keV. We have found that this peak shifts closer to 6.4 keV (i.e. higher energies) when lower frequencies are probed. This is consistent with the redder parts of the line emitted at smaller radii, and so they vary on smaller time-scales, and their light crossing time is smaller. Building an exact model however, requires a full treatment of the problem, including general relativistic effects and the interplay between the direct and reflected components in the observed data. To understand the lags, we plan to explore these details in the near future. This important result highlights what can be achieved with current telescopes in the brightest AGN NGC~4151. Larger collecting areas in future X-ray missions (e.g. LOFT and ATHENA), will make it possible to perform detailed studies in frequency, lag and energy parameter spaces in this and many other objects. 

\section*{Acknowlegement}
We Thank the referee for the useful comments that made some points clearer. AZ acknowledges support from the Department of Astronomy at UMDCP. ACF thanks the Royal society for support. CSR acknowledges support from NASA under the Astrophysics Theory Program (grant NNX10AE41G) and the Suzaku Guest Observer program (grant NNX10AR31G). This work is based on observations obtained with XMM-Newton, an ESA science mission with instruments and contributions directly funded by ESA Member States and the USA (NASA).

\bibliography{bibliography}
\label{lastpage}

\end{document}